\documentclass[aps, prl, 10pt, twocolumn,preprintnumbers,amsmath,amssymb]{revtex4}
\usepackage{amssymb}
\usepackage{amsmath}
\usepackage{graphicx}
\usepackage{color}

\begin{document}
\title{Interaction-induced photon blockade using an atomically thin mirror embedded in a microcavity}

\author{Sina Zeytino\u{g}lu}
\affiliation{Institute for Quantum Electronics, ETH Z\"urich, CH-8093 Zurich, Switzerland.}
\author{Atac \.Imamo\u{g}lu}
\affiliation{Institute for Quantum Electronics, ETH Z\"urich, CH-8093 Zurich, Switzerland.}

\date{\today }

\begin{abstract}
Narrow dark resonances associated with electromagnetically induced transparency play a key role in enhancing photon-photon interactions. The schemes realized to date relied on the existence of long-lived atomic states with strong van der Waals interactions. Here, we show that by placing an atomically thin semiconductor with ultra-fast radiative decay rate inside a \textcolor{black}{0D} cavity, it is possible to obtain narrow dark or bright resonances in transmission whose width could be much smaller than that of the cavity and bare exciton decay rates. While breaking of translational invariance places a limit on the width of the dark resonance width, it is possible to obtain a narrow bright resonance that is resilient against disorder by tuning the cavity away from the excitonic transition. Resonant excitation of this bright resonance yields strong photon antibunching even in the limit where the interaction strength is arbitrarily smaller than the non-Markovian disorder broadening and the radiative decay rate of the bare exciton. Our findings suggest that atomically thin semiconductors could pave the way for realization of strongly interacting photonic systems in the solid-state.
\end{abstract}

\maketitle

Monolayers of transition metal dichalcogenides (TMD) such as
MoSe$_2$ or WSe$_2$ constitute a new class of two dimensional (2D)
direct band-gap
semiconductors~\cite{Radisavljevic2011,Splendiani2010,Baugher2014,Britnell2013}.
Lowest energy elementary optical excitations in TMDs in the absence
of free electrons or holes are excitons with an ultra-large binding
energy of $\sim 0.5$~eV~\cite{Chernikov2014}. Remarkably, recent
experiments have demonstrated predominantly spontaneous emission
limited exciton transition linewidths in clean MoSe$_2$ or WSe$_2$
flakes that are either suspended or embedded in hexagonal boron
nitride (hBN) layers. Since radiative broadening can dominate over
disorder induced inhomogeneous broadening, clean TMD monolayers can
be considered as ideal two-dimensional (2D) optical materials,
realizing atomically thin
mirrors~\cite{Back2018,Scuri2018,Jiang2018}.

In this Letter, we show that it is possible to obtain strong photon
blockade effect~\cite{Werner1999,Verger2006,Peyronel2012,Baur2014,Pritchard2010,Petrosyan2011,Gorshkov2011,Jaksch2000} by placing a TMD monolayer inside a 2D microcavity,
even in the limit where the cavity and the exciton radiative decay rate, \textcolor{black}{as well as the disorder broadening of the exciton}
is much larger than the exciton-exciton interaction. When the
exciton is resonant with a low quality (Q) factor cavity mode, the
coupled system exhibits a dark resonance: the minimum achievable
width of this dark transmission window is
limited only by non-radiative exciton line broadening. In contrast, when the
exciton resonance is red detuned from the cavity mode, the linewidth of the
transmission \textit{peak} associated with the lower polariton can be much smaller
than the exciton disorder broadening as well as the
radiative decay rate of the bare exciton and cavity modes.
Hence, in this limit, the exciton-exciton interaction strength required to observe large
photon antibunching in cavity transmission is limited only by the -- much weaker -- Markovian dephasing or non-radiative decay processes. \textcolor{black}{Our results indicate that} particularly for dipolar excitons
with enhanced interactions, it should be possible to obtain strong photon blockade
effect.

\textcolor{black}{Before proceeding, we note that remarkable progress in realization of photon blockade has been achieved using} electromagnetically induced transparency (EIT) \cite{Peyronel2012, Baur2014}, \textcolor{black}{which}
 describes the modification of the  optical response of a medium stemming from the pumping of a driven atomic or solid-state system into a dark state~\cite{Fleischhauer2005}. Normally, the presence of an excited metastable
state that is immune to radiative decay is considered to be an
essential requirement for EIT. This metastable state and the
ground-state are coupled by coherent laser fields to a common bright
state with a large dissipation rate. The dark state that remains
uncoupled from the electromagnetic field is a superposition of the
ground and metastable states. If the atoms in the metastable state
have strong interactions, then it is possible to observe a blockade
effect where \textcolor{black}{excitation of a  second nearby atom
to its metastable state is prohibited.}
This is the essence of Rydberg blockade where the strong dipolar
interactions between atoms lead to quantum correlations between transmitted photons~\cite{Peyronel2012}.

As we argue below, the cavity-TMD system we are analyzing forms an analog of EIT if we associate the cavity mode with the bright resonance and the exciton mode with the metastable resonance. The Hamiltonian of the system is
\begin{align}
H  =  H_{\mathrm{TMD}} + H_{\mathrm{cavity}} + H_{\mathrm{int}} +
H_{\mathrm{laser}} + H_{\mathrm{bath}}
\end{align}
where
\begin{align}
H_{\mathrm{TMD}} = \sum_{k_{||}} \left[ \omega_{exc}(k_{||}) x_{k_{||}}^{\dagger} x_{k_{||}}  +
U_{x-x} \sum_{k'_{||},q} x_{k_{||}+q}^{\dagger}  x_{k'_{||}-q}^{\dagger} x_{k'_{||}} x_{k_{||}}\right]
,
\end{align}
\begin{align}
H_{\mathrm{cav}} =  \omega_c a_c^{\dagger} a_c ,
\end{align}
\begin{align}
H_{\mathrm{bath}} = \sum_{\mathbf{k}} \omega_{\mathbf{k}} b_{\mathbf{k}}^{\dagger} b_{\mathbf{k}} +  \sum_{\mathbf{k}}
\left[ \xi_{\mathbf{k}} a_c^{\dagger} b_{\mathbf{k}} + h.c. \right] ,
\end{align}
\begin{align}
H_{\mathrm{int}} = \sum_{k_{||}} \left[ g_c F^*(k_{||}) x_{k_{||}}^{\dagger} a_c + h.c.
\right] , \label{eq:intHam}
\end{align}
\begin{align}
H_{\mathrm{laser}} = \left[ \Omega_0 a_c  + h.c. \right] .
\end{align}
Here, we assumed exciton coupling to a single zero dimensional (0D) \textcolor{black}{fundamental} cavity mode $a_c$ in a structure where the photonic confinement along the $z$-direction is much stronger than the lateral confinement. To simplify the expressions, we set $\hbar = 1$ and expressed frequencies in a frame rotating with the incident optical frequency $\omega_L$. As a consequence, $\omega_{exc}(k_{||}) \rightarrow \omega_{exc}(k_{||}=0) - \omega_L + k_{||}^2/(2m_{exc})$ and $\omega_{c}
\rightarrow \omega_{c} - \omega_L$; here, $k_{||}$ denotes the in-plane momentum of the exciton. The exciton-exciton interaction
is described as a contact interaction with strength $U_{x-x}$; this
is justified in the low density limit of interest even for dipolar
2D excitons. To describe the coupling between the excitons and the cavity mode,
we used the definition $a_c = \sum_{k_{||}} F(k_{||}) a_{k_{||}}$, where $F(k_{||})$ is the
fourier transform of the cavity mode function in the plane of the TMD flake, and $a_{k_{||}}$ are the annihilation operators for the 2D cavity field modes of momentum $k_{||}$.  By integrating out the cavity coupling to free-space vacuum modes $b_{\mathbf{k}}$ described by $H_{\mathrm{bath}}$ in Markov approximation, we obtain the
Heisenberg equations of motion that includes the cavity decay at
rate $\kappa_c\equiv \xi^2 \rho(\omega_c)$, where $\rho(\omega_c)$ is the density of
states of the free space radiation modes, as well as the associated noise terms. $\Omega_0$ is the coupling strength between the coherent probe laser and the cavity field.

The assumed form of exciton-cavity coupling and the
absence of direct coupling of excitons to free space vacuum modes is central to
our \textcolor{black}{analogy between the cavity TMD system and the EIT setup}. To justify this form, we first recall that due to
conservation of in-plane momentum in a translationally invariant 2D
cavity-exciton system, each exciton mode with in-plane momentum $k$ \cite{ Janne2013, Bettles2016, Zeytinoglu2017, Shahmoon2017}
couples exclusively to a single 2D cavity mode with identical momentum
with strength $g_c = \sqrt{ \Gamma_{rad} c / L_z}$~\footnote{For
practical cavities confinement along the $z$-direction which underlies our argument is achieved
only for $k_{||} < k_{||,cut-off}$}. Here $\Gamma_{rad}$ is the spontaneous
emission rate of excitons in free-space, $c$ is the speed of light
and $L_z$ is the length of the 2D cavity along the direction
orthogonal to the monolayer plane. The case where a 0D cavity mode couples
to 2D excitons can also be described approximately by $g_c$, if the in-plane mode confinement is weak such that  $F(k_{||})$ can be approximated by a delta function $\delta_{k_{||},0}$.
In this regime, the eigenmodes of the coupled system will consist of lower and upper polariton modes that
are split by an energy $\simeq 2 g_c$. We emphasize
that translationally invariant excitons embedded in a cavity
acquire a finite decay rate exclusively due to their coherent
coupling to the cavity \textcolor{black}{with} a finite mirror loss rate
$\kappa_c$. Moreover, fast excitonic radiative decay $\Gamma_{rad}$ in free-space
emerges as an advantage, since large $\Gamma_{rad}$ leads to
enhanced coherent coupling $g_c$.

\begin{figure}
\centering
\includegraphics[width= \linewidth, trim= {0 0 10cm 0}, clip]{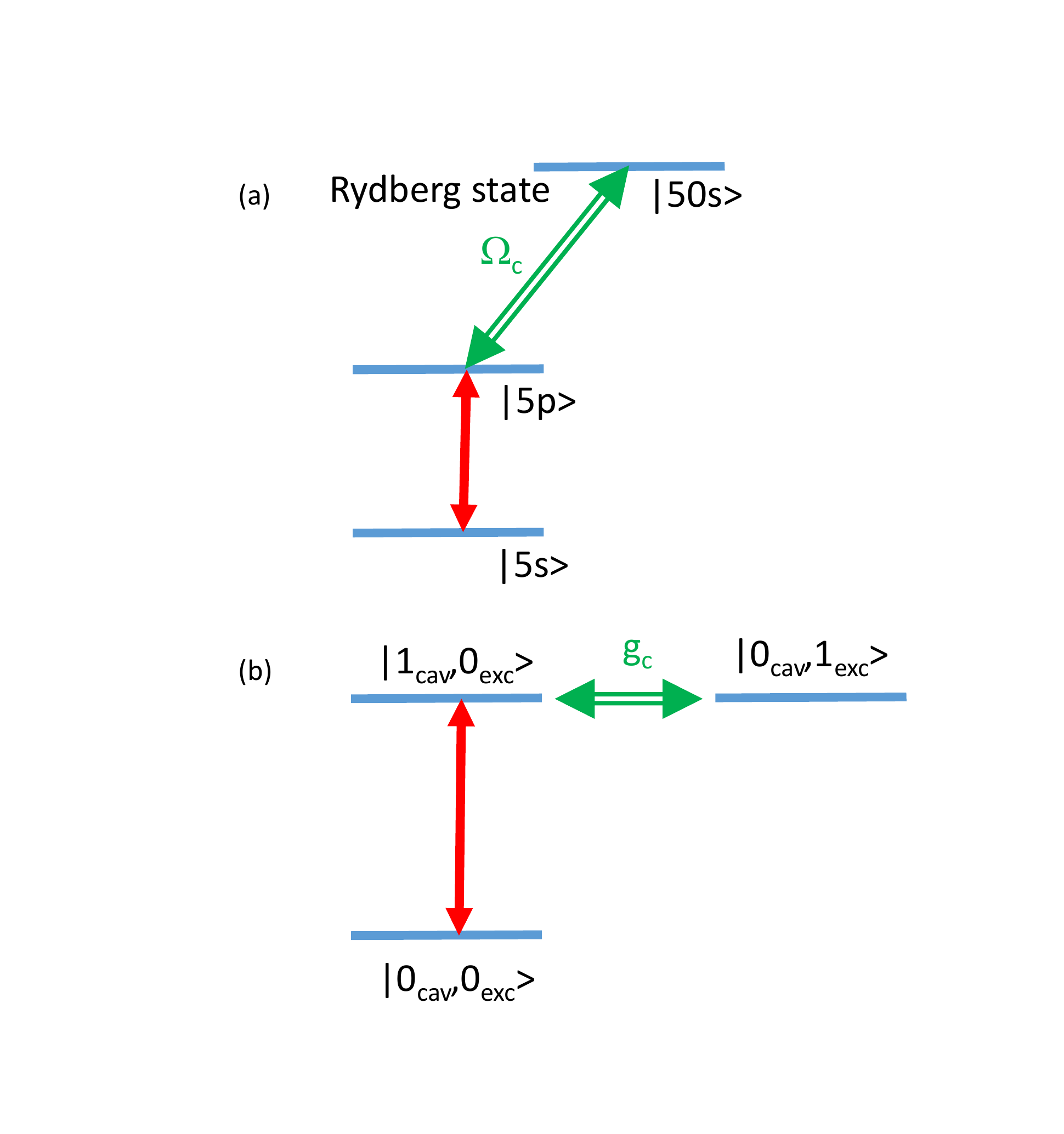}
\caption{The summary of the analogy between the cavity- TMD and conventional Rydberg EIT schemes. The role of the
fast decaying $5p$ state of the Rydberg atom is fulfilled by the radiatively
broadened cavity mode excitation, while the counterpart of the metastable Rydberg state is the excitonic transition of the TMD. Finally, the coherent drive in the Rydberg EIT scheme is replaced by the coherent coupling of the excitonic
transition of the TMD to the cavity mode. }
\label{fig:Analogy}
\end{figure}

The analogy between the cavity-TMD system and the
conventional EIT setup employed with Rydberg atoms is a direct
consequence of vanishing direct radiative decay of the exciton mode. As
indicated in Figure~ $\ref{fig:Analogy}$, in the cavity-TMD scheme,
the role of collective excitation from the ground level to the first
excited p-level in Rydberg-EIT is replaced by the cavity mode
excitation. The counterpart of the coherent laser coupling of the
p-level to the metastable Rydberg state is the vacuum-field coupling
of the cavity mode to the TMD exciton.The EIT condition in the
Rydberg scheme is achieved by creation of a superposition excitation
of the ground and Rydberg states that suppresses light scattering
from the intermediate p-level. In the cavity-TMD scheme, the
corresponding dark state is a coherent superposition of the ground
state with an excitonic excitation with vanishing cavity-mode
amplitude
\begin{align}
| \Psi \rangle  \simeq ( \alpha + \beta x_0^{\dagger} ) | 0, G \rangle,
\label{eq:DarkState}
\end{align}
where $|0\rangle$ and $|G\rangle$ denote the vacuum state of the
cavity and the TMD, respectively.  Expression in Eq.
($\ref{eq:DarkState}$) is the steady-state of the coupled system in
the limit of weak drive, provided that the incident drive laser and
the bare exciton transition are on resonance [$\omega_{exc}(k=0) =
0$]. Upon formation of this coherent superposition, the cavity mode
occupancy and consequently cavity transmission vanishes and the
incident field experiences perfect reflection. On the other hand,
when the drive laser is resonant with the polaritonic transitions in
the cavity-TMD system, the transmission spectrum of the system
exhibits bright resonance peaks.

Keeping the correspondence with Rydberg blockade, we envision two
scenarios: in the first case, we assume $\omega_{exc}(k=0) =
\omega_{c}$ and $\kappa_c > 4 g_c$ where the coupled system exhibits no polariton splitting but a
dark resonance in transmission. Following the EIT analogy, we find
that the width of the transmission dip on resonance is given by
$g_c^2/\kappa_c$. If the TMD excitons are subject to non-radiative
decay $\gamma_{nr}$ or pure dephasing $\gamma_{deph}$ stemming from
coupling to Markovian reservoirs, the condition for the observation
of a dark resonance is given by $g_c^2/\kappa_c > \gamma_d =
\gamma_{nr} + \gamma_{deph}$. Observation of quantum correlations
between transmitted photons in this regime would in turn require
$U_{x-x} \simeq g_0^2/\kappa_c > \gamma_d$.  This simple analysis indicates that strong photon antibunching is observable even when $\Gamma_{rad} \gg U_{x-x}$, provided Eq.(8) is satisfied.

Recent experimental observations suggest that $\gamma_d$ plays a relatively minor role in comparison to disorder scattering in determining the non-radiative line broadening of TMD excitons \cite{Back2018, Selig2016}. Therefore, the resonant case that we have just described will in practice be limited by disorder scattering that leads to a
strongly energy-dependent, non-Markovian line broadening for low
momentum excitons. The non-Markovian nature of the
disorder induced decay rate follows from the fact that scattering
processes due to disorder conserve energy, and the phase space
available for the exciton to scatter into can be strongly energy
dependent. The disorder induced decay rate is determined by the imaginary part of the corresponding self energy $\Im[\Sigma_{\mathrm{dis}}(\omega)]$ (see Supplementary Material). The frequency window in which $\Im[\Sigma_{\mathrm{dis}}(\omega)]$ is non-zero
is typically of the same order of magnitude as its maximum: we denote the latter as $\delta_{\mathrm{dis}}$. When $g_c^2/\kappa_c \le \delta_{dis}$, the effects of
disorder can be approximated by that of an effective Markovian
reservoir. In the opposite limit $g_c^2/\kappa_c > \delta_{dis}$,
disorder has a vanishing effect on the transmission; the asymmetry
of the polariton transmission peaks stemming from quantum
interference can be observed in this regime.

The suppression of non-Markovian line broadening in the limit where coherent coupling
exceeds Doppler broadening was already highlighted in the context of
EIT assisted sum frequency conversion in Doppler broadened atomic
gases~\cite{Harris1990}. Similarly, strongly non-Markovian nature of disorder broadening
ensures that the cavity-exciton system exhibits steep dispersion
when the dark-resonance width exceeds the disorder broadening. As an
important consequence, exciton-exciton interactions can render the
system anharmonic even in the limit where interaction
strength is weaker than the exciton decay rate.  Nevertheless, we find that strong photon antibunching in this
limit can only be observed in the limit $U_{x-x} \ge max \left[\delta_{dis},g_c^2/\kappa_c\right]$.

\begin{figure}
\includegraphics[width= \linewidth]{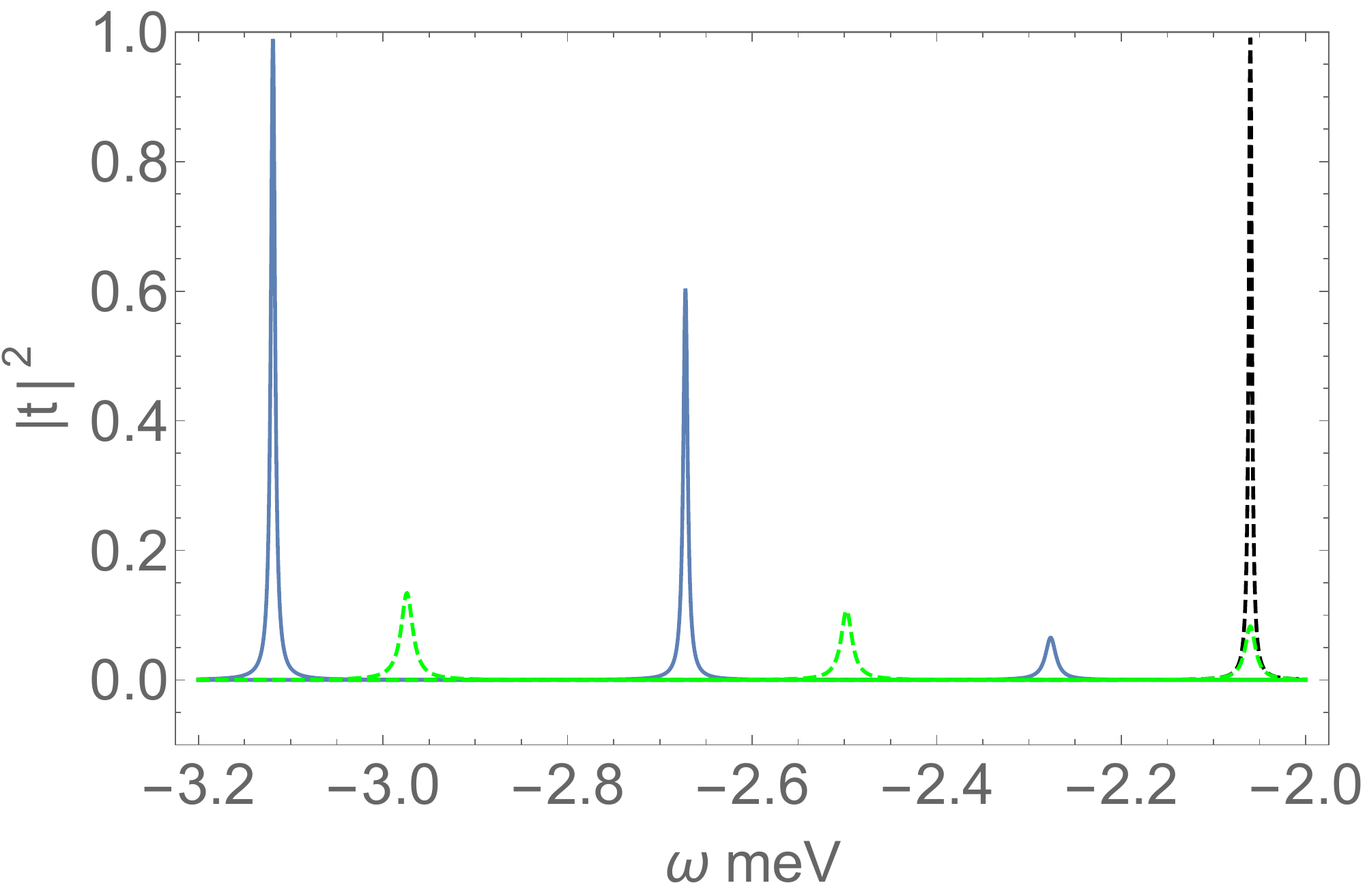}
\caption{ Transmision $|t|^2$ spectrum of the lower polariton for $\kappa_c~=~0.2$ meV ,
$\Delta_c = 100$ meV for increasing values of the cavity coupling $g_c = 14.5 - 17.5$ meV, with 1.5 meV
intervals.  The blue solid traces take account of the decay of the $k=0$ exciton due to disorder treated
with correlated coherent potential approximation (CPA), where $\delta_{\mathrm{dis}}=1$ meV, while the dashed green traces only consider Markovian dephasing whose rate is set to $\gamma_d=\delta_{\mathrm{dis}}/100=0.01$~meV. The black dashed trace is meant to serve as a reference for the transmission of the lower polariton for $g_c = 14.5$~meV, if neither Markovian dephasing nor disorder induced decay is taken into account.  As the $g_c$ is increased such that the $g_c^2/\Delta_c \gg \delta_{\mathrm{dis}}$, the adverse effects of the disorder
induced decay become negligible, while the effect of Markovian dephasing persists. The peak transmission is given by $\left | \frac{\Gamma_{LP} }{\Gamma_{LP}  + \delta_{\mathrm{dis}}}\right |^2$ for the case of Markovian dephasing. The difference in the transition frequency of the exciton with Markovian broadening and non-Markovian broadening is due to the real part
of the disorder induced self energy (see Supplementary material).}
\label{fig:Trans}
\end{figure}

\begin{figure}
\includegraphics[width= \linewidth]{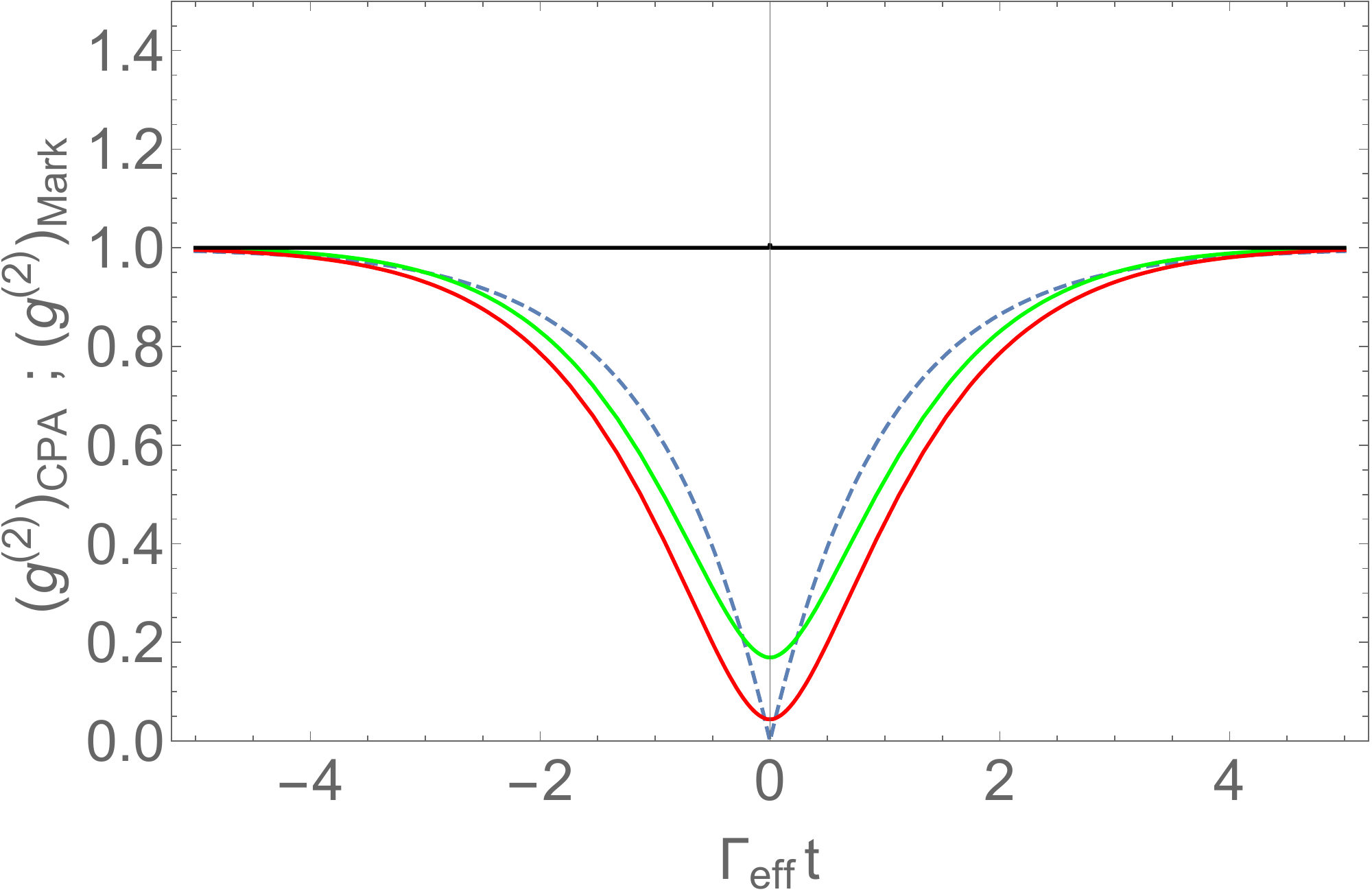}
\caption{ Time dependent second order correllation function $g^{(2)}(t)$ when the incoming photons are resonant with the lower polariton
in the case of CPA disorder (in red and green )  and Markovian dephasing (in black). $U = 0.02 $ meV (in red) $U=0.01$ meV (in Green),
$g_c = 20 $ meV , $\kappa_c = 0.1$ meV , $\Delta_c = 100$ meV. Disorder self energy is calculated such that $\delta_{\mathrm{dis}} = 1$ meV, and $E_c \equiv \frac{\hbar^2 \eta^{-2}}{2 m} = \sigma $, where $\eta$ is the correlation length of the disorder potential and $\sigma$ is the
variance of the disorder potential (see Supplementary Material).
For these parameters and interaction strength $U = 0.02$ meV, the Markovian dephasing
with $\gamma_d = \delta_{\mathrm{dis}}$ completely destroys the transmission peak
of the lower polariton and results in a flat $g^{(2)}(0)\approx1$, implying Poissonian statistics.
The blue dashed line is to guide the eye and represents the exponential decay with the radiative decay rate ($\Gamma_{LP}$) of the lower polariton.}
\label{fig:g2}
\end{figure}

The overcome this limitation, we consider a second scenario where we
assume a large detuning between the cavity and exciton resonances [$
\Delta_c = \omega_c - \omega_{exc}(k=0) > g_c$]. In this limit, the
coupled system exhibits a narrow bright resonance red detuned from
the bare exciton resonance by $g_c^2/\Delta_c$. This case is
analogous to Rydberg blockade experiments in the regime where the
incident photons are detuned from the intermediate state. \textcolor{black}{However,}
due to the \textcolor{black}{strong} non-Markovian character of the disorder broadening, we
find that it is possible to completely suppress the adverse effects
$\Im[\Sigma_{\mathrm{dis}}(\omega)]$. In Fig. $\ref{fig:Trans}$, we plot
the transmission spectrum of the lower polariton for different values of $g_c$
from $14.5-17.5$ meV with 1.5 meV intervals, and observe the recovery of transmission at the lower polariton resonance as the detuning of the lower polariton from the bare exciton
exceeds $\delta_{\mathrm{dis}}$ (i.e., $g_c^2/\Delta_c \gg\delta_{\mathrm{dis}}$). As a result, we obtain strong photon antibunching even in
the limit where $U_{x-x} < \delta_{\mathrm{dis}}, g_c^2/\kappa_c$, as long as the
detuning of the narrow bright resonance is much larger than the
strength of the disorder $g_c^2/\Delta_c \gg \delta_{\mathrm{dis}}$
and $U_{x-x} > max \left[ \Gamma_{LP} = \kappa_c g_c^2/\Delta_c^2 , \gamma_d \right]$. In stark contrast, Markovian processes that lead to the same exciton line broadening as disorder scattering would have lead to Poissonian statistics of the transmitted light [Fig.~$\ref{fig:g2}$].

The calculate the photon
correlation function $g^{(2)}(t)$, we use the scattering matrix approach presented
in Ref.~\cite{Shi2015}, which is reviewed in the Supplementary
Material. The conventional wisdom \cite{Fleischhauer2005}
suggests that probe photons injected at an energy where the
transmission has the sharpest features result in the largest {\sl
amplification of the interaction effects} in photon correlations.
When the probe laser is tuned on resonance with the sharp lower
polariton transmission feature, injection of the first photon into
the cavity-exciton system will shift the resonance by $\approx U_{x-x}$.
Since the conditional probability that the successive photons will
be transmitted (reflected) is reduced (enhanced) in this limit, we
expect to see strong photon antibunching (bunching) in $g^{(2)}(t)$.

For the $g^{(2)}(t)$ calculation depicted in Fig.~\ref{fig:g2} we choose $\Delta_c = 100$~meV, $g_c = 20$ meV, $\delta_{dis} = 1$ meV and $U_{x-x}\simeq 10 - 20
\, \mu eV$. The experimentally reported values of $g_c$ range from $10$~meV to more than $40$~meV, depending on the employed cavity structure. Recent experiments demonstrating TMD monolayers as atomically thin mirrors indicate that in clean samples disorder broadening can indeed be as narrow as $0.5$~meV and possibly lower. The principal unknown parameter is $U_{x-x}$: the value we chose was motivated by recently measured interaction strength of GaAs excitons confined to  $A = 2 \mu$m$^2$~\cite{Rodriguez2017, Munoz2017}. While a detailed calculation taking into account non-local screening effect \cite{Berkelbach2013,Cudazzo2011} has not been carried out for $U_{x-x}$, we expect TMD exciton interaction strength to be comparable to that in GaAs.

To further enhance $U_{x-x}$ it is desirable to use a heterobilayer
structure where an intra-layer exciton couples resonantly to an
inter-layer (indirect) exciton by coherent electron or hole
tunneling (J); such structures have been implemented in GaAs
structures to realize dipolar polaritons~\cite{Cristofolini2012} with enhanced interactions~\cite{Togan2018,Rosenberg2018}.
In the limit where the indirect exciton is tuned into resonance with
the bright resonance and $g_c^2/\Delta_c > J > \Delta_{dis}$ is
satisfied, it would be possible to obtain a bright resonance with a
permanent dipole moment.

In the  $g^{(2)}(t)$ calculations depicted in Fig.~3 (green and red curves), we take into account radiative decay and disorder scattering, but neglect line broadening of excitons
stemming from coupling to additional reservoirs ($\gamma_d$). Long-wavelength phonon coupling between
high and low momentum intra-valley excitons, as well as relaxation
of bright intra-valley excitons into inter-valley dark exciton
states by short-wavelength phonon emission could lead to $\gamma_d > 0$ and limit the
minimum achievable linewidth of the bright polariton resonance. We
emphasize however, that due to the non-Markovian character of the
phonon bath at ultra-low temperatures, strong exciton-cavity coupling could strongly suppress
both of these channels; in particular, by choosing $g_c^2/\Delta_c $
to be comparable to the electron-hole exchange interaction, it is
possible to eliminate relaxation into dark exciton states by short
wavelength phonons. Another potential decay channel for high
momentum excitons which will modify the imaginary part of exciton
self-energy is radiative coupling to guided modes which in turn have
a finite lifetime due to the finite sample size. Coupling to the
guided modes result in a frequency independent (Markovian)
contribution to the excitonic self energy; the latter decay channel
may be suppressed by using in-plane photonic band-gap structures eliminating
guided modes that are resonant with the lower-polariton mode.

In summary, we show that photon blockade regime can be achieved in a
cavity-TMD system even when the excition-exciton interaction
strength is much smaller than the cavity and exciton radiative decay
rates. The resilience of quantum correlations to disorder scattering stems from the  non-Markovian nature of the associated exciton coherence decay. Remarkably, the only fundamental requirement for observation
of strong photon antibunching is $U_{x-x} > \gamma_d$. Given the immense possibilities for controlling the excitonic properties of TMD monolayers using electrical gates or structured dielectric environment, we expect the demonstration of photon blockade to establish cavity-TMD system as a building block of strongly correlated photonic systems.

We acknowledge useful discussions with Aymeric Delteil, Martin Kroner, Misha Lukin, Li-Bing Tan, Dominik Wild and Susanne Yelin. This work is supported by a European Research Council (ERC) Advanced investigator grant (POLTDES).

\end{document}